\documentclass{aa}  
\newcommand{\bb}{$\langle B \rangle$}
\newcommand{\bbm}{\langle B \rangle}
\newcommand{\vsini}{$v$ sin $i$}
\newcommand{\teff}{$\text{T}_{\text{eff}}$}

\newcommand{\vmic}{$v_{\textit{mic}}$}
\newcommand{\vmac}{$v_{\textit{mac}}$}
\newcommand{\logg}{log $g$}

\usepackage{graphicx}
\usepackage{txfonts}
\begin{document} 

   \title{Magnetic fields of intermediate mass T Tauri stars}


\author{
  A. Lavail\inst{1,2}
  \and
  O. Kochukhov\inst{1}
  \and
  G. A. J. Hussain\inst{2}
  \and
  E. Alecian\inst{3}
  \and
  G. J. Herczeg\inst{4}
  \and
  C. Johns-Krull\inst{5}
}

\institute{
  Department of Physics and Astronomy, Uppsala University, Box 516, SE-751 20 Uppsala, Sweden\\
  \email{alexis.lavail@physics.uu.se}
  \and
  ESO, Karl-Schwarzschild-Strasse 2, 85748 Garching, Germany
  \and
  UJF-Grenoble 1 / CNRS-INSU, Institut de Planétologie et d’Astrophysique de Grenoble (IPAG) UMR
  5274, Grenoble, F-38041, France
  \and
  The Kavli Institute for Astronomy and Astrophysics, Peking University, Beijing 100871, China
  \and
  Department of Physics and Astronomy, Rice University, 6100 Main Street, Houston, TX 77005, USA
}

  \date{Received September 4, 2017; accepted November 13, 2017}

 
	\abstract
{}
{
    In this paper, we aim to measure the strength of the surface magnetic fields for a sample of five intermediate mass T Tauri stars and one low mass T Tauri star from late-F to mid-K spectral types. While magnetic fields of T Tauri stars at the low mass range have been extensively characterized, our work complements previous studies towards the intermediate mass range; this complementary study is key to evaluate how magnetic fields evolve during the transition from a convective to a radiative core.
}
{
	We studied the Zeeman broadening of magnetically sensitive spectral lines in the H-band spectra obtained with the CRIRES high-resolution near-infrared spectrometer. These data are modelled using magnetic spectral synthesis and model atmospheres. Additional constraints on non-magnetic line broadening mechanisms are obtained from modelling molecular lines in the K band or atomic lines in the optical wavelength region.
}
{
	We detect and measure mean surface magnetic fields for five of the six stars in our sample: CHXR 28, COUP 107, V2062 Oph, V1149 Sco, and Par 2441. Magnetic field strengths inferred from the most magnetically sensitive diagnostic line range from 0.8 to 1.8 kG. We also estimate a magnetic field strength of 1.9 kG for COUP 107 from an alternative diagnostic. The magnetic field on YLW 19 is the weakest in our sample and is marginally detected, with a strength of 0.8 kG.
}
{We populate an uncharted area of the pre-main-sequence HR diagram with mean magnetic field measurements from high-resolution near-infrared spectra. Our sample of intermediate mass T Tauri stars in general exhibits weaker magnetic fields than their lower mass counterparts. Our measurements will be used in combination with other spectropolarimetric studies of intermediate mass and lower mass T Tauri stars to provide input into pre-main-sequence stellar evolutionary models.}

\keywords{
	stars: pre-main sequence --
	stars: magnetic field --
	line: profiles
}

   \maketitle
%
\section{Introduction}
Magnetic fields influence physical processes in stars at all stages of their evolution. In the pre-main-sequence (PMS) phase in particular magnetic fields govern accretion processes (see \citealt{2007prpl.conf..479B,2016ARA&A..54..135H} for a review of magnetospheric accretion in classical T Tauri stars), which affect stellar formation and angular momentum evolution. However, the evolution of stellar magnetic fields from the birth line, when all stars are fully convective, and along the PMS, when stars may develop a radiative core, is still not thoroughly understood.

Intermediate mass T Tauri stars (IMTTS; defined here to have mass $1 \text{M}_\sun \leqslant \text{M} \leqslant 4 \text{M}_\sun$) are key targets to study the magnetic behaviour of PMS stars, test theories of magnetic field formation, and understand the incidence and properties of magnetic fields in more evolved stars. These IMTTS are the precursors to the PMS Herbig Ae/Be stars and eventually to the A/B type stars on the main sequence. Less than 10\% of these stars host a magnetic field \citep{2007pms..conf...89P,2009ARA&A..47..333D}, which is generally strong, of simple structure (dipole-like), and stable over at least several decades.

A recent large spectropolarimetric survey aimed at studying the incidence of magnetism among the precursors of A- and B-type stars, Herbig Ae/Be stars, revealed similar properties \citep{2013MNRAS.429.1001A}, indicating that the global magnetic field is likely conserved between the PMS and main-sequence phases, and therefore must be produced at an earlier phase of evolution. As IMTTS are precursors of the Herbig Ae/Be stars, the study of their magnetic fields can shed new light on the origin of magnetic fields in intermediate mass stars. 

Magnetic fields in T Tauri stars have been investigated either by measuring the magnetic broadening of the intensity profiles of infrared \ion{Ti}{i} lines around 22000 {\AA} (\citealt{2007ApJ...664..975J} -- hereafter JK07; \citealt{2008AJ....136.2286Y}, and \citealt{2011ApJ...729...83Y}) or using the Zeeman Doppler imaging modelling of circular polarization spectra (\citealt{2007MNRAS.380.1297D}, \citealt{2008MNRAS.386.1234D}, \citealt{2009MNRAS.398..189H}, reviewed in \citealt{2012AN....333....4H}).
The former method is the only method that is capable of providing an unbiased estimate of the total magnetic flux since polarimetric observables are severely affected by cancellation of opposite polarities at small spatial scales. For their sample of 14 classical T Tauri stars, JK07 measured mean magnetic field strengths from 1.12 kG to 2.90 kG and found no significant correlation between the observed field strengths and field strengths predicted by simple magnetospheric accretion models. \citet{2011ApJ...729...83Y} measured field strengths ranging from 1.30 kG to 3.45 kG in their sample of 14 additional T Tauri stars and reported that there is no correlation between field strengths {\bb} and stellar age for the entire sample of 33 T Tauri stars studied using magnetic line broadening. However, they found that magnetic fluxes $ 4 \pi R_{*}^2 \bbm$ decrease steadily with stellar age.

In this paper, we report magnetic field strengths for a sample of six additional T Tauri stars obtained using magnetic line broadening measurements. Our targets are mostly in the intermediate mass range, with masses ranging from 1.1 $\text{M}_{\sun}$ to 2.0 $\text{M}_{\sun}$, which has not been investigated as well as the lower mass range of the classical T Tauri star regime. One target (CHXR 28) is however in the low mass end at 0.6 $\text{M}_{\sun}$. Considering the range of spectral types of our sample, well-established cool stars magnetic diagnostics such as \ion{Ti}{I} lines around 22200~{\AA} and \ion{FeH}{} lines around 10000~{\AA} are unsuitable. We instead use \ion{Fe}{i} lines around 15650~{\AA} -- used by \citet{1995ApJ...439..939V} to analyse the spectra of the active K2V star $\epsilon$ Eridani -- which are deep enough in our spectra.

This paper is organized as follows. In Sect. \ref{section:observations}, we discuss the acquisition and reduction of CRIRES spectra in the near-infrared H and K bands used in our study. We describe our methods to estimate projected rotational velocities {\vsini}, analyse the Zeeman broadening of magnetically sensitive spectral lines, and estimate mean magnetic field modulus {\bb} in Sect. \ref{section:methods}. We report our measurements for each star in the sample in Sect. \ref{section:results} and discuss our results in Sect. \ref{section:conclusion}.
%
\section{Observations and data reduction}
\label{section:observations}
\begin{figure}
	\centering
	\includegraphics[width=8.8cm]{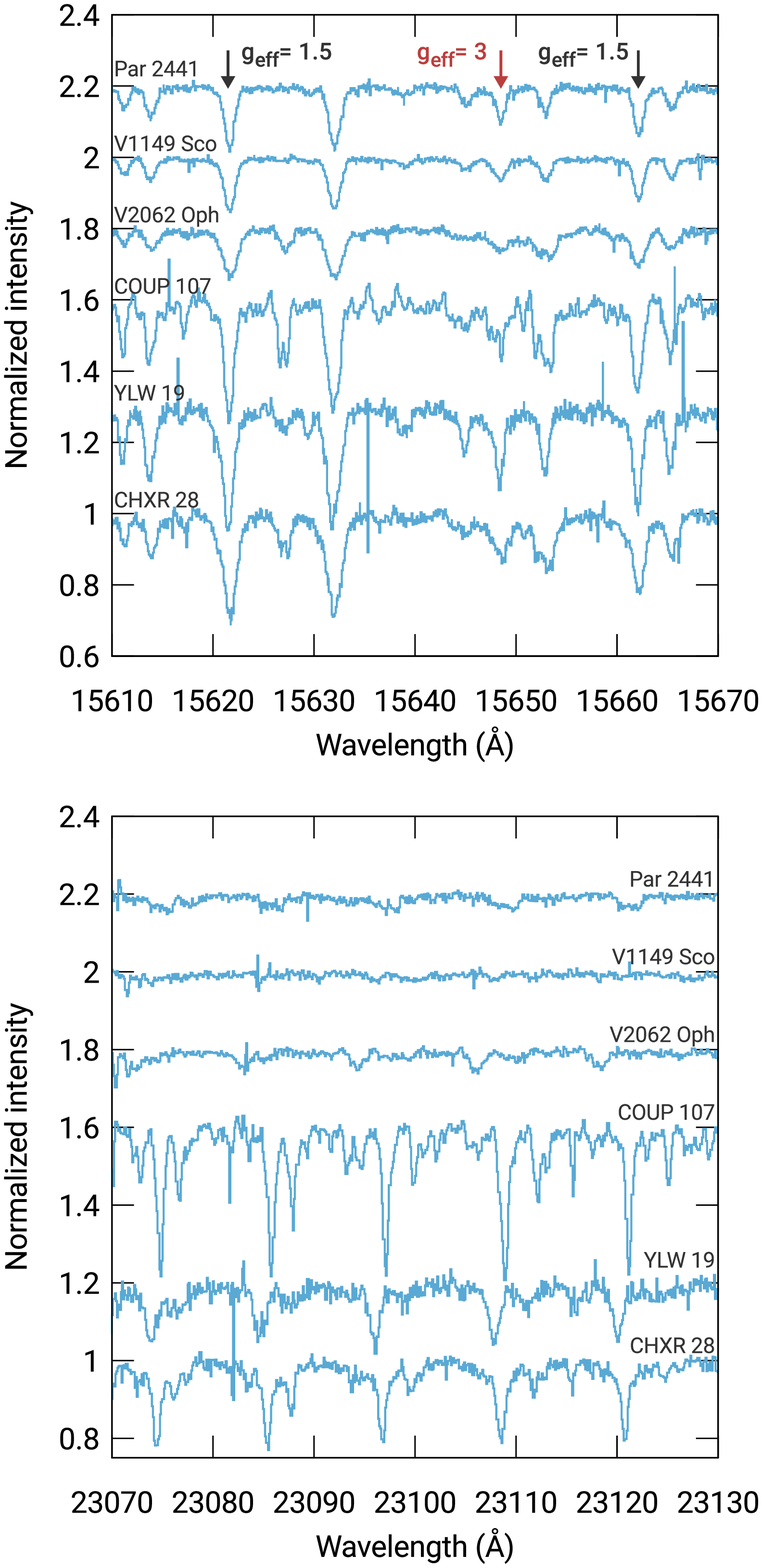}
	\caption{ Normalized CRIRES spectra in the H band (upper panel)
		and the K band (lower panel). The spectra are shifted
		vertically with stellar effective temperature rising
		upwards. The arrows in the upper panel indicate the spectral lines used
		in the magnetic field analysis. The central arrow
		indicates the magnetically sensitive spectral line
		\ion{Fe}{i} 15648.5 {\AA} with $g_{\text{eff}} = 3$.
   }
   \label{figure:HK}
\end{figure}
%
We observed six stars with the high-resolution near-infrared CRIRES spectrograph \citep{2004SPIE.5492.1218K} mounted at the Very Large Telescope  to obtain high-resolution, $R \approx 10^5$ , spectra in the H and K band. The journal of our observations is presented in Table~\ref{table:observations}. The CRIRES adaptive optics correction improved the seeing to $0.25\arcsec$ in the H band and $0.16\arcsec$ in the K band.  The stars were placed in the $0.2\arcsec$ slit, which yields a spectral resolution of $R = 10^5$.  The dispersed light was recorded by four $512\times1024$ detectors, producing four distinct spectra that each cover $\sim 1500$ km s$^{-1}$ and are separated by $\sim 400$ km s$^{-1}$. The data were obtained in ABBA nods to improve background correction.

Data were reduced with custom-written codes in IDL. The AB and, separately, the BA pairs were subtracted from each other first. Bad pixels were eliminated using a cosmic ray rejection routine. The spectra for each exposure were then obtained by summing the number of counts over a $\pm0.5 \arcsec$ ($\pm6$ pixels) extraction region. Nearby regions on the detector were used to subtract any residual flux. The odd-even effect on detectors 1 and 4 was corrected on each individual spectrum.  The four spectra were then individually normalized.  The total spectrum was calculated from the median of each pixel whenever multiple exposures of the same target in a given wavelength region were available. The wavelength scale was calculated using telluric absorption lines and had an rms precision of $0.05-0.2$ km s$^{-1}$ depending on the segment. Telluric absorption was corrected with observations of A-type stars immediately following our science observations. The wavelength scale was then shifted to the heliocentric velocity frame. The reduced spectra are shown in Fig.~\ref{figure:HK}.

CHXR 28 is part of a triple system with a tight companion (Ab separated by $0.142 \arcsec$) and a wide companion (B) located $1.85 \arcsec$ from the tight subsystem \citep{2008ApJ...683..844L}. The A and B components are resolved and the slit was aligned along the position angle between the two components.  The spectra of both components were resolved on the detector and were extracted separately. Both components were first fit with Gaussian profiles at each pixel to minimize contamination. The Gaussian profile of one component was then subtracted off the detector, leaving the remaining component. We discarded the spectra of the cooler secondary component CHX 10B and studied only the primary CHX 10A (CHXR 28). V1149 Sco was observed twice on the same night in both H and K bands. We co-added the individual spectra to improve the signal-to-noise ratio (S/N) to more than 200 in the H-band setting.

%
\begin{table*}
\caption{Journal of observations}
\label{table:observations}
\centering
\begin{tabular}{l c c c c c c }     
\hline\hline
Target & Other name & UT date & UT time at start & Wavelength setting & Integration time (s) & S/N \\
\hline
            &			&				&			& 			& & \\
\object{CHXR 28}		& CHX 10a	& 2012-08-10	& 00:00:12	& 1574.4 nm	& $720 $& 84 \\
            &			& 2012-08-10	& 00:16:48	& 2325.2 nm	& $960 $& 67\\
            &			&				&			& 			& & \\

\object{YLW 19}		& ISO-Oph 1	& 2012-07-31	& 03:11:41	& 1574.4 nm & $1200 $& 70\\
            &			& 2012-07-31	& 03:36:04	& 2325.2 nm & $1200 $& 55\\
            &			&				&			&			& & \\

\object{COUP 107}	& KM Ori	& 2012-12-26	& 01:47:23	& 1574.4 nm	& $1200 $& 86\\
            &			& 2012-12-26	& 02:12:54	& 2325.0 nm	& $1200 $& 48\\
            &			&				& 			& 			& & \\

\object{V2062 Oph}   & ROX 44	& 2012-05-01	& 05:03:01	& 1574.4 nm	& $720 $& 147\\
            &			& 2012-05-01	& 05:19:37	& 2325.2 nm	& $960 $ & 146\\
                        &			&				& 			& 			& & \\

\object{V1149 Sco}	& HD 143006	& 2012-05-01	& 04:04:11 	& 1574.4 nm	& $720 $& 156\\
            &			& 2012-05-01	& 06:47:56 	& 1574.4 nm & $720 $& 169\\
            &			& 2012-05-01	& 04:19:47 	& 2325.2 nm & $720 $& 199\\
            &			& 2012-05-01	& 07:04:07 	& 2325.2 nm & $720 $& 151\\
            &			&				& 			& 			& & \\

\object{Par 2441}	& HD 294260	& 2012-11-08	& 06:32:56	& 1574.4 nm	& $1200 $& 145\\
			&			& 2012-11-08	& 06:57:30	& 2325.2 nm & $1200 $& 133\\
                        &			&				& 			& 			& & \\
\hline
\end{tabular}
\end{table*}
%
\section{Magnetic analysis methods}
\label{section:methods}
\subsection{Basic principles}
Our analysis relies on the comparison between observed and synthetic intensity profiles of absorption lines in the near-infrared H and K bands (around 15600 {\AA} and 23000 {\AA, respectively}). Specifically, our aim is to detect the magnetic broadening -- also called Zeeman broadening -- of spectral lines. The presence of a magnetic field splits the lower and upper levels of atomic transitions according to their Land\'e factors and J quantum numbers, leading to the splitting of spectral lines into groups of $\sigma$ and $\pi$ components. In cool stars, owing to the combination of magnetic field strengths and projected rotational velocities involved, this splitting tends to result in additional line broadening rather than in overall line splitting. The additional width of the Zeeman-broadened spectral line is proportional to the strength of the magnetic field, to the square of the wavelength, and to the effective Land\'e factor of the spectral line, according to the following formula:
$$ \Delta \lambda_{\text{B}} = 4.67 \times 10^{-7} g_\text{eff} {{\lambda}_0}^2 \langle B \rangle $$
where the wavelength of the centre of the line ${\lambda}_0$ is in~{\AA}, the magnetic field modulus averaged over the stellar surface {\bb} is in kG, and the effective Landé factor $g_\text{eff}$ is unitless.

The magnetic broadening is sensitive to the magnetic fields at all spatial scales that affect the formation of photospheric lines. However, the geometry of the magnetic field cannot be meaningfully recovered unless the Zeeman components are fully resolved. As a consequence, a homogeneous purely radial field -- or a simple distribution of field strengths -- is commonly assumed by the magnetic broadening analyses; see \citet{2014IAUS..302...25H} for a comparison of magnetic broadening and ZDI methods applied to PMS stars. 

The effective Land\'e factor represents the sensitivity of the spectral line to the magnetic field and measures the wavelength separation between the centre of gravity of $\sigma$ components and the line centre \citep{1982SoPh...77..285L}. In our spectral synthesis we used detailed Zeeman splitting patterns for each transition. When available, these splitting patterns were derived from the Land\'e factors of upper and lower atomic levels listed in the {\tt VALD3} \citep{vald2015} database. In other cases, we computed Land\'e factors under the assumption of LS coupling.

To obtain a more robust estimate of the magnetic field, it is advantageous to measure Zeeman broadening at long wavelengths, typically in the near-infrared, and of magnetically sensitive spectral lines, with a high effective Land\'e factor. It is the most efficient method to reliably estimate magnetic field strength in cool stars and in IMTTS.
%
\subsection{Magnetic spectral synthesis}
In our analysis, we assume solar abundance and adopted a microturbulence (\vmic) of $2 \text{ km } \text{s}^{-1}$ while neglecting macroturbulence, and accounted for the instrumental $R = 100000$ broadening. Furthermore, a simple magnetic field configuration is assumed in which the magnetic field is purely radial, with constant strength over the entire stellar surface. Alternatively, if the spectral lines exhibited complex shapes, in particular the $g_{\text{eff}} = 3$ \ion{Fe}{i} line at 15648.5~\AA, we adopted a two-component model in  which a fraction of the stellar surface is covered with a magnetic field of constant strength, while the rest is  non-magnetic. To create synthetic spectra, we adopted values of the effective temperature from the literature (see Table \ref{table:parameters}) and assumed a {\logg} value of $3.5$. We used a grid of {\tt MARCS} model atmospheres \citep{marcs2008}, which have a step in effective temperature of 250 K. We chose the grid models with the closest {\teff} to our targets. The synthetic spectra have been generated via the magnetic spectral synthesis code {\tt SYNMAST} \citep{OK2007pms} which numerically treats the polarized radiative transfer for a list of atomic and molecular lines in a given input model atmosphere. Line blending and individual Zeeman patterns are treated in detail.
%
\begin{table}
\caption{Magnetic diagnostics with effective Land\'e factor and oscillator strengths used in this work}
\label{table:diagnostics}
\centering
\begin{tabular}{c c c }     
\hline\hline
Wavelength	& log $gf$ & $g_{\text{eff}}$ \\
({\AA})			& &  \\
\hline
            &			&						\\
15621.654	&	+0.59		&	1.5					\\
15648.510	&	-0.68	&	3.0					\\
15662.013 	&	+0.37		&	1.5					\\
            &			&						\\
\hline
\end{tabular}
\end{table}
%
\subsection{Improvement of oscillator strengths}
The first step of our analysis consisted of improving the quality of the oscillator strengths obtained from {\tt VALD3}. Corrections to the oscillator strength values change the depth of spectral lines, but not their widths. Therefore, the magnetic field strength {\bb} values that we derive from the Zeeman broadening measurements, which depend primarily on the spectral line widths, are largely unaffected. However, incorrect oscillator strengths can affect our results if we use moderately blended lines, in case there is a relative oscillator strength error between the main component and the blending line.

\begin{figure*}
	\centering
	\includegraphics[width=18.0cm]{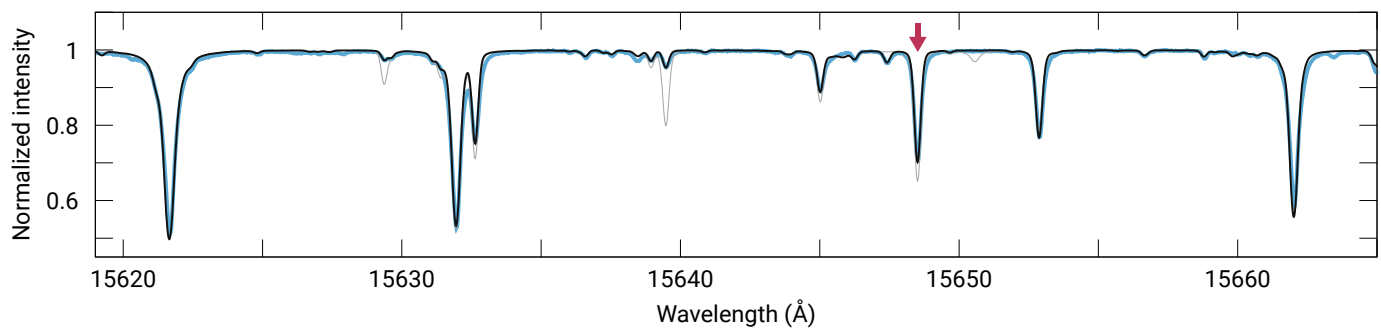}
	\caption{
		Observed solar spectrum in the H band from \cite{1991aass.book.....L} (blue thick line) compared with synthetic spectra generated with the original oscillator strengths from VALD3 (grey thin line) and corrected oscillator strengths (black thick line). The spectral line indicated with the red arrow is the high Landé factor line \ion {Fe}{i} 15648.5 {\AA}  used in this study.}
	\label{figure:loggf}
\end{figure*}
%
To correct oscillator strengths, we computed a synthetic solar spectrum for the wavelength region covered by our CRIRES observations in the H band; we used the same method as that employed when producing synthetic spectra for our targets via the MARCS model atmosphere with solar parameters and compared this solar synthetic spectrum to the solar observations from \cite{1991aass.book.....L}. We then used {\tt BINMAG4}\footnote{http://www.astro.uu.se/\textasciitilde{}oleg/binmag.html} to adjust the oscillator strength of individual lines automatically, by fitting the synthetic spectrum to the observations via $\chi ^2$ minimization. The improvement in the fit of the synthetic spectrum to the observed solar spectrum is shown in Fig. \ref{figure:loggf}.
%
\begin{table*}
\caption{Parameters of the studied stars}
\label{table:parameters}
\centering
\begin{tabular}{c c c c c c}
\hline\hline
Stars	& Spectral type	& {\teff}	& {\vsini}  					& Mass  		& References  	\\
		&				& (K)		& ($\text{ km } \text{s}^{-1}$)	& ($M_{\sun}$)	&				\\
\hline 
CHXR 28		& K7		& 4060		& $ 7.5 \pm 1.1	$	& $0.60$		& H97, D13  \\[5pt]
YLW 19		& K4		& 4590      & $ 10.5 \pm 1.1$	& $1.10$		& E11   \\[5pt]
COUP 107	& K0-K4		& 4730      & $ 6.5 \pm 1.1	$	& $1.40$		& H97, DR12  \\[5pt]
V2062 Oph	& K3		& 4730      & $ 17.5 \pm 1.0$	& $1.25$ 		& BA92, W16   \\[5pt]
V1149 Sco	& G5-G8		& 5875      & $ 11.5 \pm 1.9$	& $1.15$ 		& S06, G07, T07, LM12   \\[5pt]
Par 2441	& F9		& 6115      & $ 13.0 \pm 1.2$	& $2.00$ 		& C04  \\[5pt]
\hline
\end{tabular}
\tablefoot{The stellar masses have been estimated with the luminosities and effective temperatures from the litterature, using the Yale–Potsdam Stellar Isochrones grid \citep{2017ApJ...838..161S}.}
\tablebib{
	(BA92)~\citet{bouvier1992};
    (C04)~\citet{2004AJ....128.1294C};
    (D13)~\citet{daemgen2013};
    (DR12)~\citet{2012ApJ...748...14D};
    (E11)~\citet{erickson2011}; 
    (G07)~\citet{2007A&A...461..183G};
    (H97)~\citet{1997AJ....113.1733H};
    (LM12)~\cite{2012ApJ...758...31L};
	(S06)~\citet{2006A&A...458..173S}; 
    (T07)~\citet{2007ApJS..168..297T}
    (W16)~\citet{willson2016}.
}
\end{table*}
\subsection{Determination of mean field modulus}
One line in our wavelength range is very magnetically sensitive; i.e. the \ion{Fe}{i} line at 15648.5~{\AA}, with $g_\text{eff} = 3$. However, this line can either be blended or too weak for some of our targets to yield a reliable estimate of the magnetic field strength. Therefore we also used two stronger lines, which are less magnetically sensitive; i.e. the \ion{Fe}{i} lines at 15622~{\AA} and 15662~{\AA} with $g_\text{eff} = 1.5$. These features still provide diagnostics as sensitive as the best diagnostics in the optical. Table \ref{table:diagnostics} lists the spectral lines used as diagnostics in this paper, together with their effective Landé factors $g_\text{eff}$ and oscillator strengths.

\begin{figure}
	\centering
	\includegraphics[width=8.8cm]{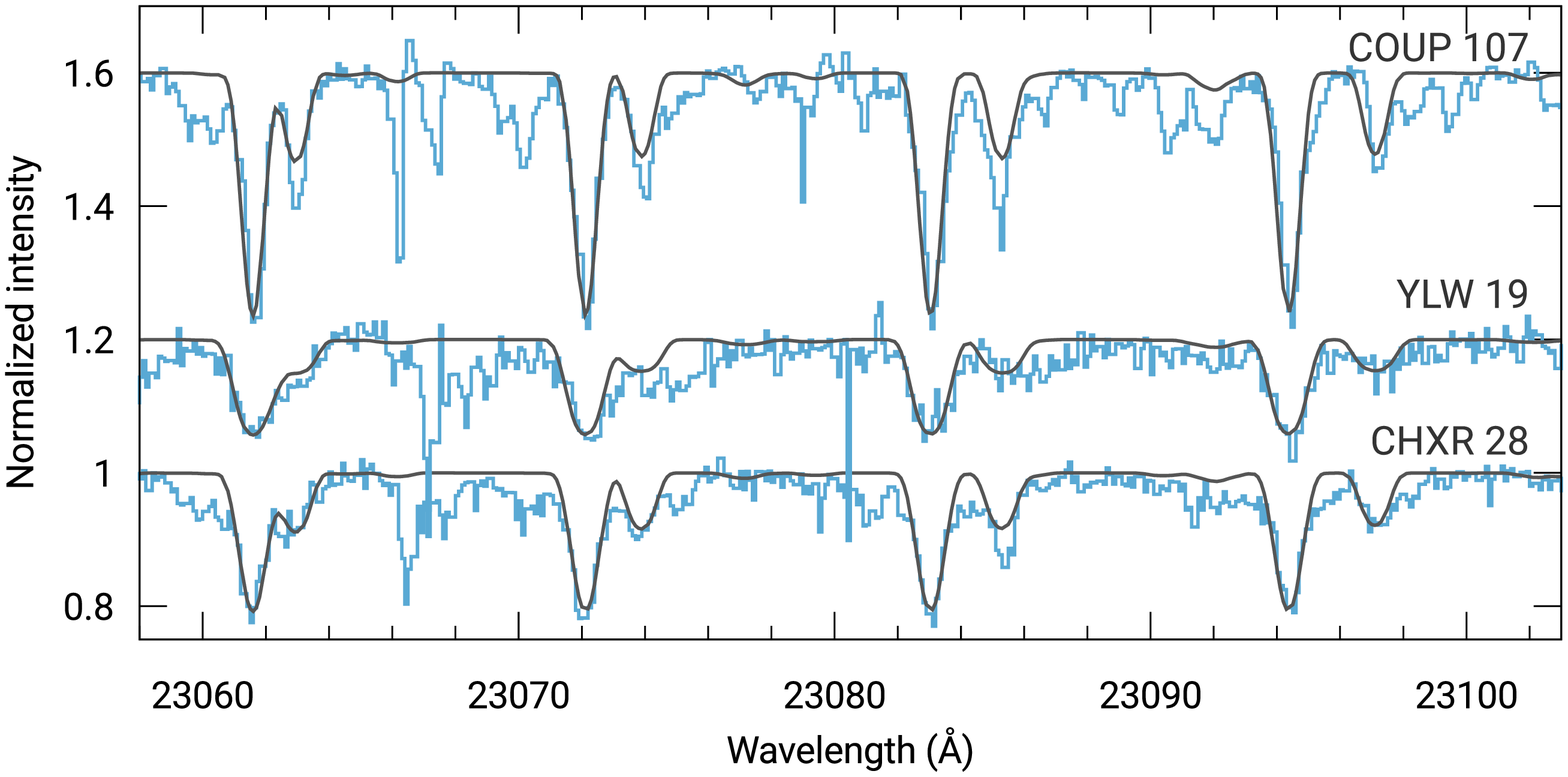}
	\caption{
		Observed and synthetic spectra of COUP 107, YLW 19, and CHXR 28 in the K band with the molecular CO lines used for the determination of {\vsini}.
	}
    \label{fig:vsini-ir}
\end{figure}

\begin{figure}
	\centering
	\includegraphics[width=8.8cm]{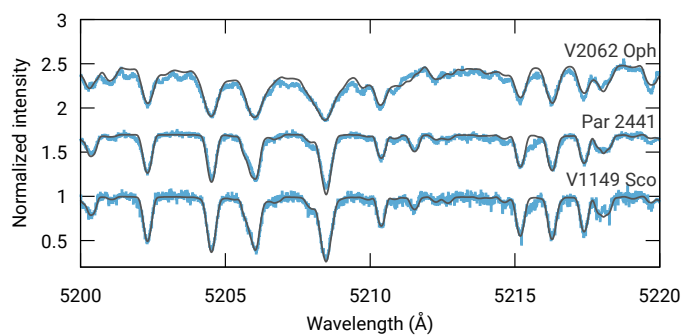}
	\caption{
		Observed and synthetic spectra of V2062 Oph, Par 2441, and V1149 Sco in the optical domain.
	}
    \label{fig:vsini-optical} 
\end{figure}

To constrain non-magnetic broadening (e.g.  $v$sin$i$)  of the cooler targets (CHXR 28, YLW 19, and COUP 107), we used magnetically insensitive CO lines in the K band, typically the strongest CO lines at 23072~{\AA}, 23083~{\AA}, and 23094~{\AA}. The hotter targets (V2062 Oph, V1149 Sco, and Par 2441) do not have strong enough spectral features in this K-band wavelength setting. Instead, for V1149 Sco and Par 2441, we used strong optical unblended \ion{Fe}{i} lines with low $g_\text{eff}$ values in $R = 65000$ archival {ESPaDOnS} spectra retrieved from the PolarBase database \citep{2014PASP..126..469P}. The {ESPaDOnS} instrument is a high-resolution spectropolarimeter mounted on the 3.6 m Canada-France-Hawaii Telescope. For V2062 Oph, the same strong optical unblended \ion{Fe}{i} lines were modelled in a $R = 66000$ archival spectrum\footnote{The UVES spectrum was retrieved through the ESO Science Archive Facility} from the UVES echelle spectrograph \citep{2000SPIE.4008..534D} mounted at the ESO UT2 Very Large Telescope. We adopted fixed values for non-magnetic broadening parameters other than for {\vsini}, where \vmic $= 2 \text{ km } \text{s}^{-1}$ and \vmac $= 0 \text{ km } \text{s}^{-1}$. The values of {\vsini} and their associated 1$\sigma$ uncertainties were determined through $\chi ^2$ minimization. These results are presented in Table \ref{table:parameters}. The best fits to the K-band CRIRES spectra of CHXR 28, YLW 19, and COUP 107 are shown in Fig. \ref{fig:vsini-ir}. The best fits to the optical spectra of V1149 Sco, Par 2441, and V2062 Oph are shown in Fig. \ref{fig:vsini-optical}.
%

For each of our targets, we produced a grid of synthetic spectra using our corrected oscillator strengths, where the strength of the magnetic field is assumed to be purely radial and covering the entire stellar surface, and is allowed to vary between 0.0 kG to 6.0 kG in increments of 0.1 kG. For each value of {\bb} we compared the synthetic spectrum with the observed spectrum in narrow windows centred on the three spectral lines of interest: \ion{Fe}{i} 15648.5~{\AA}, \ion{Fe}{i} 15622~{\AA}, and \ion{Fe}{i} 15662~{\AA}.
However, in the case of CHXR 28 and V1149 Sco, the \ion{Fe}{I} line at 15648.5~{\AA} has a complex, approximately triangular shape, which is not reproduced well with a simple one-component magnetic model, where the entire surface of the star is covered with an uniform magnetic field strength. For these two targets, we used a two-component magnetic model, which covers a given fraction of the surface with a constant magnetic field strength and the rest with a null magnetic field. We allowed two parameters of this model to vary while fitting the spectral lines: the strength of the magnetic field and fraction of the stellar surface covered with magnetic field, also called the filling factor. We adopted the filling factor and field strength corresponding to the best fit of the \ion{Fe}{I} line at 15648.5~{\AA}.
%
\begin{figure}
	\centering
	\includegraphics[width=8.8cm]{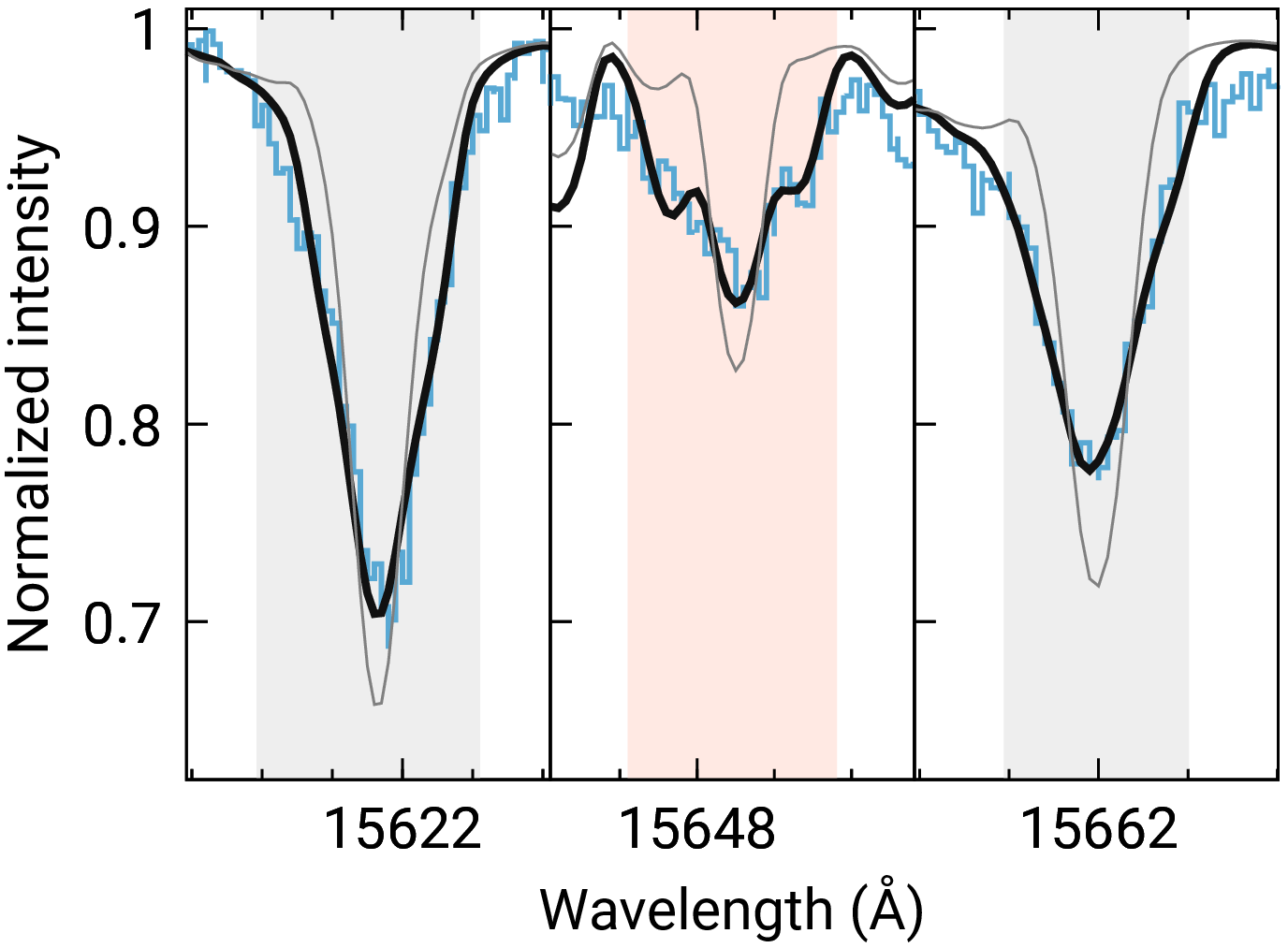}
	\caption{ 
    	Comparison of the observed spectra (blue histogram), the magnetic synthetic spectra, yielding the best fit to the observations (black thick line) and the non-magnetic synthetic spectra (grey thin line) for the three lines of interest \ion{Fe}{i} 15622~{\AA} (left), \ion{Fe}{i} 15648.5~{\AA} (middle), and \ion{Fe}{i} 15662~{\AA} (right) of CHXR 28.
	}
	\label{figure:ylw19_fit}
\end{figure}
%

\begin{figure}
	\centering
	\includegraphics[width=8.8cm]{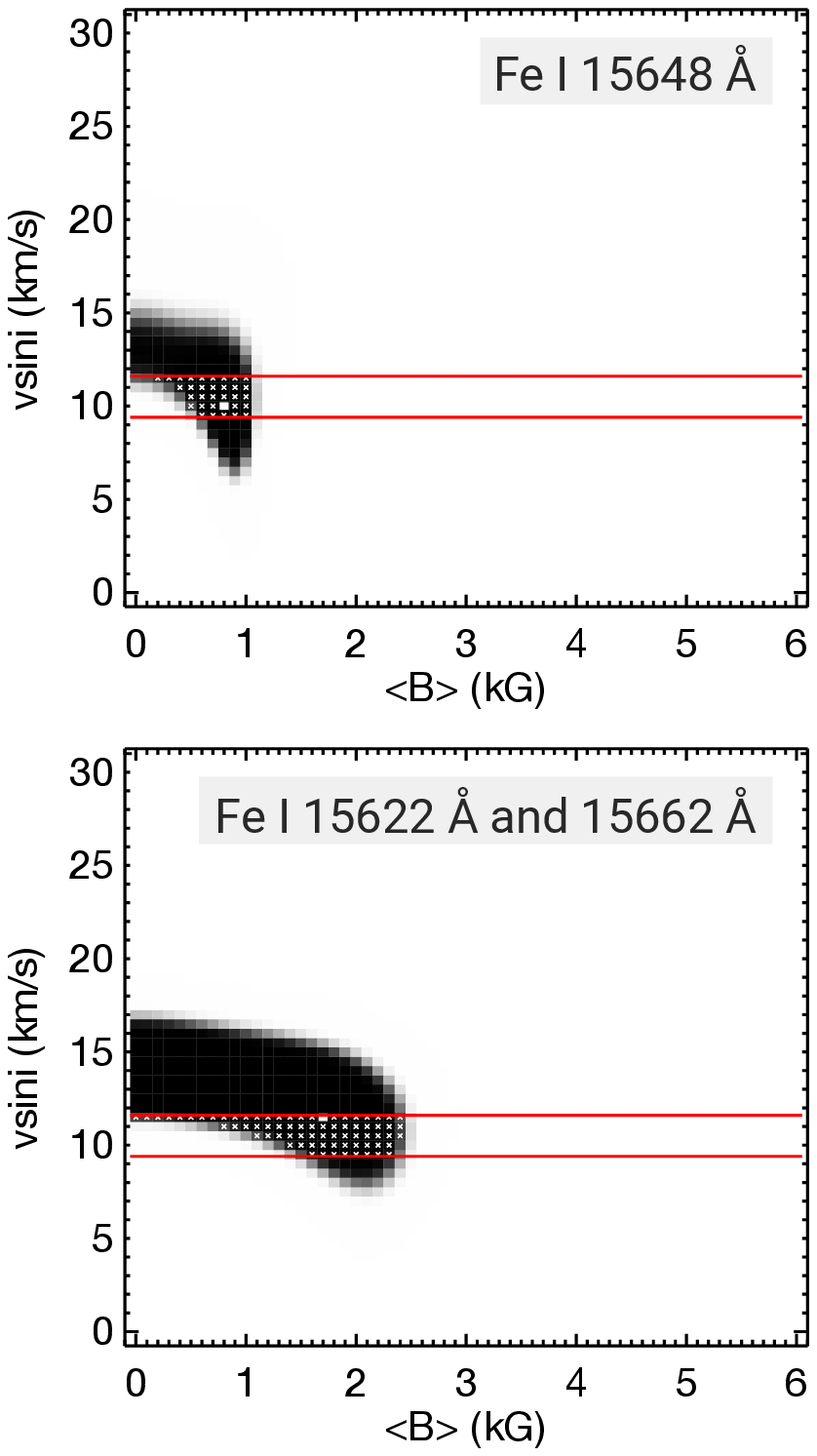}
	\caption{
	 $\chi ^2$ map for the three lines of interest. \ion{Fe}{i} 15648.5~{\AA} (top), \ion{Fe}{i} 15622~{\AA} and \ion{Fe}{i} 15662 {\AA} (bottom) for YLW 19 are shown. The small white rectangle indicates the minimum $\chi ^2$ and the white crosses indicate values within 1$\sigma$ from the minimum and within the 1$\sigma$ limits on {\vsini} indicated by the horizontal red lines.
     \label{fig:ylw19_contour}
  	}
\end{figure}
%
T Tauri stars commonly exhibit veiling, or an excess of continuum intensity, causing the spectral lines to be shallower. To account for the veiling of spectral lines, it is customary to apply a scaling factor, or a veiling factor, to normalized synthetic spectra, which has the effect of reducing the depth of spectral lines with respect to the continuum level. We let the veiling factor vary for each set of spectral lines and for each couple of {\bb} and {\vsini}, adopting the value leading to the minimal $\chi ^2$. In other words, in this study we only use information from line shapes and do not rely in any way on the strength of spectral lines, which can also be affected by the magnetic field through the magnetic intensification mechanism. Figure~\ref{figure:ylw19_fit} illustrates how the observed spectral lines of CHXR 28 can be fitted well with a magnetic synthetic spectra. With no magnetic field, the spectral lines are too narrow and do not reproduce the shape of observed line profiles.

The same method was used to compute the $\chi ^2$ and confidence intervals for the one-component and two-component magnetic models, following \citet{2013A&A...554A..61K}. First, we computed a grid of $\chi ^2$ by fitting the synthetic spectra to the observations as a function of {\vsini} and \bb. We only retained grid points with $\chi ^2$ values within 1$\sigma$ from the minimum $\chi ^2$ , which have {\vsini} values within the independently determined lower and upper limits, as illustrated by Fig.~\ref{fig:ylw19_contour}. The difference between the highly magnetically sensitive ($g_{\text{eff}}$ = 3) \ion{Fe}{i} line at 15648.5~{\AA} and the two $g_{\text{eff}}$ = 1.5 \ion{Fe}{i} lines is clear; the latter has a weak dependence on {\bb}, therefore putting weaker constraints on the field strength for a given {\vsini}. 

\section{Results}
\label{section:results}
\subsection{CHXR 28}
\label{subsection:CHXR28}
%
\begin{figure}
	\centering
	\includegraphics[width=8.8cm]{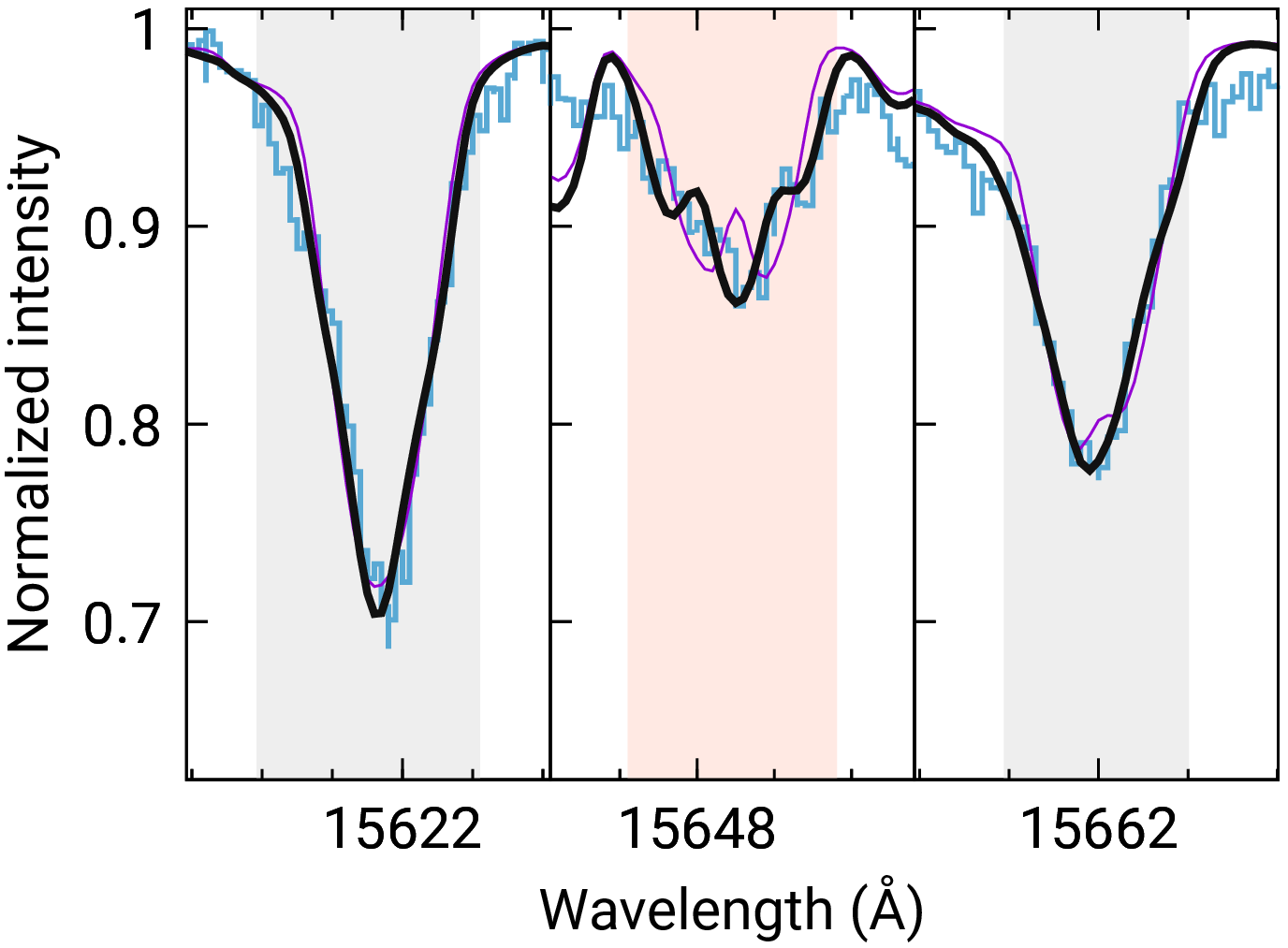}
	\caption{
	 Comparison of the best magnetic fit using a two-component model (thick black line) and a one-component model (purple thin line) to the observed spectrum of CHXR 28 (blue histogram).
     \label{fig:chx10-ff-noff}
  	}
\end{figure}
%
CHXR 28 (CHX 10A) has been identified as a member of the Cha I star-forming region \citep{luhman2004,luhman2007}. The H-band spectrum of CHXR~28 shows a complex-shaped \ion{Fe}{I} line at 15648.5~{\AA}, leading us to use a two-component magnetic model. The two-component model performs significantly better at reproducing the shape of the \ion{Fe}{I} line at 15648.5~{\AA}. This line is best reproduced with a filling factor of 60\% and {\bb} $= 1.5$ kG and has lower and upper 1$\sigma$ limits at 1.3 and 1.7 kG,  respectively. The filling factor approach lowers  the reduced ${\chi}_{r}^2$ to 1.1 from 1.3 for a single-component fit. The strong $g_{\text{eff}}$ = 1.5 \ion{Fe}{i} lines yield higher values of {\bb}. Fig.~\ref{fig:chx10-ff-noff} illustrates how the two-component model improves the fitting of the observed spectral lines in the spectrum of CHXR 28 compared to the single-component model.
The magnetic field derived for CHXR 28 from the $g_{\text{eff}}$ = 1.5 lines is systematically stronger. This result appears to be largely driven by the width of the Fe I 15662~{\AA}. On the other hand, the 15622~{\AA} line exhibits a somewhat narrower core than this strong-field model suggests. In this and other cases, we put a higher weight on the $g_{\text{eff}} = 3$ line analysis results, which are deemed to be more reliable. For this particular target we also cannot exclude an influence of the unresolved close visual companion Ab \citep[$\Delta Ks = 0.40 \pm 0.05$, $\Delta H = 0.40 \pm 0.04$;][]{2008ApJ...683..844L}.
%
\subsection{YLW 19}
\label{subsection:YLW19}
YLW 19 is a member of the $\rho$ Ophiuci cloud that is attributed to a K4 spectral type \citep{erickson2011} by optical spectroscopy; this source has a companion at $0.42\arcsec $ \citep{2003ApJ...591.1064B,2005A&A...437..611R}.
The spectral lines show little sign of magnetic broadening. The $g_{\text{eff}}$ = 3 \ion{Fe}{i} line restricts {\bb} with 1$\sigma$ between 0.2 and 1.0 kG with a best fit for {\bb} $= 0.8$ kG.
%
\subsection{COUP 107}
\label{subsection:coup107}
COUP 107 lies in the Orion nebula cluster and has a spectral type between K0 and K4 \citep{2012ApJ...748...14D,1997AJ....113.1733H}. The $g_{\text{eff}}$ = 3 \ion{Fe}{I} line at 15648.5 {\AA} is severely blended and unsuitable for our analysis in the H-band spectrum, leaving only the couple of $g_{\text{eff}}$ = 1.5 \ion{Fe}{i} lines. These two lines alone provide a weaker constraint on the magnetic field modulus, but still yield a lower and upper 1$\sigma$ limit at 1.7 and 2.2 kG, respectively, with a best-fit at \bb~$ = 1.9 $~kG.
%
\subsection{V2062 Oph}
\label{subsection:v2062oph}
V2062 Oph is a K3 star and X-ray source belonging to the $\rho$ Ophiuci cloud \citep{bouvier1992}. The $g_{\text{eff}}$ = 3 \ion{Fe}{I} line is best described with synthetic spectrum computed for {\bb} $= 1.8$ kG. The corresponding 1$\sigma$ confidence limits are 1.4 and 2.2 kG.   
%
\subsection{V1149 Sco}
\label{subsection:v1149sco}
V1149 Sco is a known member of the Upper Scorpius association with a G7 spectral type \citep{1988mcts.book.....H,2012ApJ...758...31L}. The shape of the \ion{Fe}{I} 15648.5~{\AA} line was too complex to be correctly reproduced using a single-component magnetic model, so we used a two-component model. The best fit of the \ion{Fe}{I} 15648.5~{\AA} was obtained using a filling factor of 50\% and {\bb} $= 1.4$ kG. The corresponding 1$\sigma$ limits are 1.1 and 1.7 kG. The ${\chi}_{r}^2$ value of the fit is 1.5 for a two-component model and 2.0 for a single-component model.
%
\subsection{Par 2441}
\label{subsection:par2441}
Par 2441 is a F9 star from the Ori OB1c association around the Orion nebula cluster \citep{2004AJ....128.1294C}.
The H-band spectrum of this star exhibits a marginal evidence of magnetic broadening and the fitting of the \ion{Fe}{I} line at 15648.5~{\AA} suggests {\bb} between 0.5 and 1.0 kG with a best fitting value of {\bb} $= 0.8$ kG.
\begin{table*}
\caption{Magnetic field measurement results.}	
\label{table:mf}      							
\centering										
\begin{tabular}{c c c c c c c c c}				
\hline\hline									
   Star 		&~&	  	& \bb$_{3}$	&		&	& 	 	& \bb$_{1.5}$	&		\\ 
   				& &		&  (kG)		&		&	&		&  (kG)			&		\\
   				&~& min & best 		& max	&	& min	& best			& max	\\ 
\hline											
   CHXR 28  	&~& 1.3 & 1.5 & 1.7 &  & 2.2 & 2.4 & 2.7	\\    
   YLW 19  		&~& 0.2 & 0.8 & 1.0 & & 0.0 & 1.7 & 2.4	\\
   COUP 107 	&~&  -  &  -  &  -  & & 1.7 & 1.9 & 2.2	\\
   V2062 Oph	&~& 1.4 & 1.8 & 2.2 & & 1.2 & 2.7 &	3.6	\\
   V1149 Sco 	&~& 1.1 & 1.4 & 1.7 & & 1.3 & 1.9 &	2.2	\\
   Par 2441 	&~& 0.5 & 0.8 & 1.0 & & 0.4 & 1.6 &	2.2	\\

\hline											
\end{tabular}
\tablefoot{The magnetic field has been estimated using 2 sets of spectral lines: \bb$_3$ with $g_{\text{eff}}$ = 3 \ion{Fe}{i} 15648.5~\AA, and \bb$_{1.5}$ with $g_{\text{eff}}$ = 1.5 \ion{Fe}{i} 15622~{\AA} and 15662~\AA. Columns 3 and 6 indicate the magnetic field values yielding the best fit to the spectral lines, while columns 2 and 5 and columns 4 and 7 respectively give the 1$\sigma$ lower and upper limits.}
\end{table*}
\section{Conclusions}
\label{section:conclusion}

\begin{figure*}
	\centering
	\includegraphics[width=18cm]{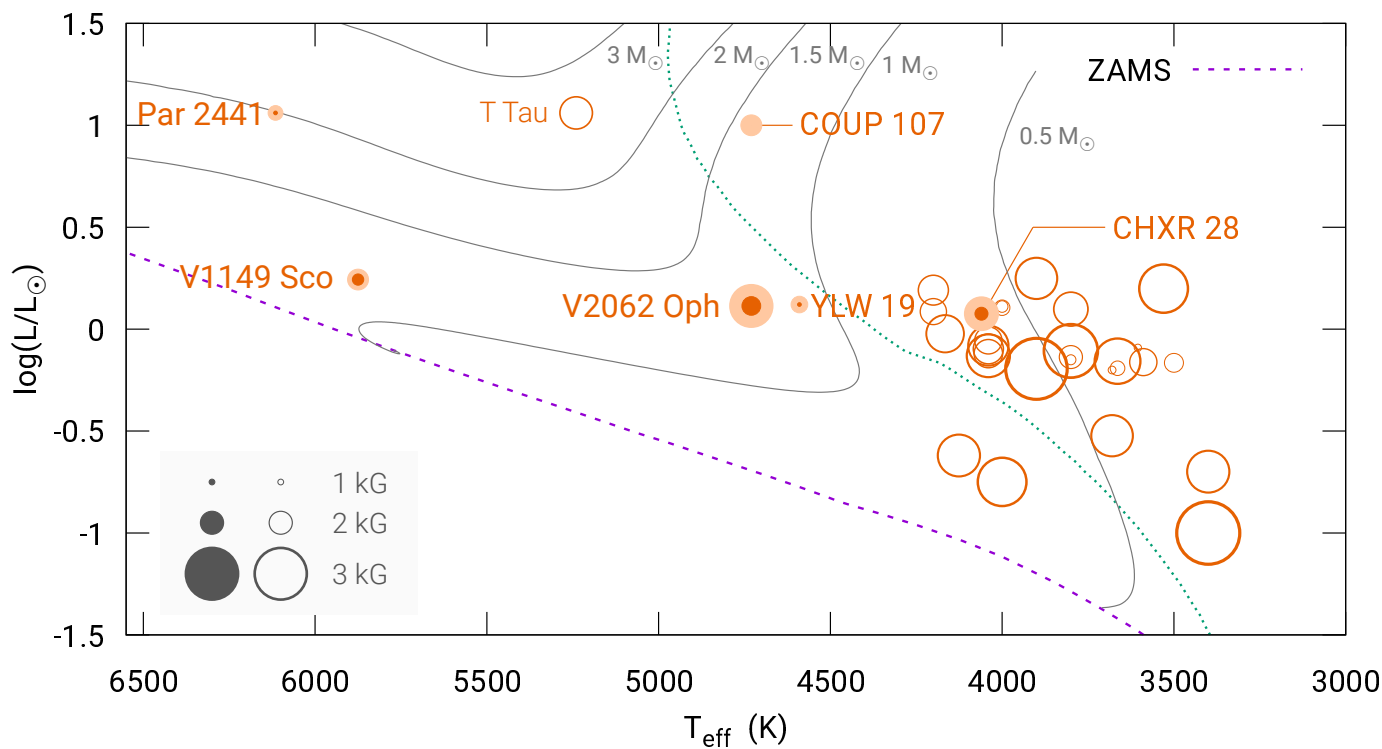}
	\caption{
		HR diagram position of T Tauri stars with magnetic field measurements from near-IR spectroscopy from this work (filled circles) and from the literature (open circles). The radius of the circles is proportional to the square of the magnetic field strength. Filled light orange and filled dark orange circles represent  \bb$_{1.5}$ and \bb$_{3}$ measurements, respectively, from Table \ref{table:mf}. We overplot the zero age main sequence (ZAMS; dashed purple line), the evolutionary tracks for 0.5, 1, 2, and 3 $\text{M}_{\sun}$ (dark lines), and the limit between fully convective and partially convective interiors (dotted green line) from the Yale–Potsdam Stellar Isochrones grid \citep{2017ApJ...838..161S}. The key in the lower left corner gives the corresponding symbol sizes for magnetic field strengths of 1, 2, and 3 kG.}
	\label{figure:hr}
\end{figure*}
%
We characterized the magnetic broadening of magnetically sensitive spectral lines for a sample of low to intermediate mass T Tauri stars and derived their average surface magnetic field strengths. We showed that \ion{Fe}{i} lines around 15600~{\AA} represent suitable magnetic diagnostics for cool stars with effective temperatures ranging from 4000 to 6000~K. Magnetic fields are detected for five out of six stars in our sample. We extended previous measurements of mean magnetic field strengths obtained with this method to mostly uncharted regions of the PMS HR diagram. We do not seem to detect magnetic fields as strong among T Tauri stars of intermediate mass as those found for their lower mass counterparts. 

Magnetic field strengths inferred from the $g_{\text{eff}}$ = 3 \ion{Fe}{i} line range from 0.8 to 1.8~kG, while magnetic field strengths inferred from the less magnetically sensitive $g_{\text{eff}}$ = 1.5 \ion{Fe}{i} lines are systematically stronger and range from 1.7 to 2.7~kG. The systematic difference between the magnetic field strength obtained with these two diagnostics hints at uncertainties in the modelling. Yet, the two sets of results are formally compatible except for the case of CHXR 28. We deem the most magnetically sensitive $g_{\text{eff}}$ = 3 diagnostic to provide the most reliable results. 

The limited size of the sample at intermediate mass does not allow us to make definitive comparison with lower mass stars. Indeed, we note that the only previous measurements of a IMTTS is 2.4 - 2.7 kG \citep[T Tau;][]{2007ApJ...664..975J,2008AJ....136.2286Y}, therefore it is important to study a larger number of targets to improve statistics. The overall comparison of magnetic field measurements in TTS of different masses is presented in Fig.~\ref{figure:hr}, which places the stars of our sample on a PMS HR diagram alongside the stars studied by \citet{2007ApJ...664..975J}, \citet{2008AJ....136.2286Y}, and \citet{2011ApJ...729...83Y}. Evolutionary tracks for stellar masses from 0.5 to 4.0 $M_\sun$, the limit between fully convective and partially radiative interiors, and the zero age main sequence from the recent Yale-Potsdam Stellar Isochrones \citep{2017ApJ...838..161S} provide an evolutionary context. Except for T Tau, stars with the strongest field lie in the fully convective zone on the right side of the HR diagram.

Ongoing spectropolarimetric studies of IMTTS, including objects from our sample (V1149 Sco, Par 2441), will enable a better understanding of the magnetic properties of these stars and will provide information on the large-scale magnetic field topologies of the global component of their magnetic fields (Villebrun et. al in prep). 

The limited simultaneous wavelength coverage of current or previously used high-resolution near-IR spectrometers is clearly the main limiting factor of this study and preceding studies by only allowing us to use one or a few spectral lines as magnetic diagnostics. While the K band magnetically sensitive \ion{Ti}{i} lines are strong in the spectra of cooler low mass T Tauri stars, these lines are not appropriate to study hotter IMTTS. Additionally, IMTTS cover a wider range of effective temperatures and it is challenging to find a single magnetically sensitive spectral line that is strong in the spectra of stars with effective temperatures ranging from roughly 4000 to more than 6000 K. We expect that both recently commissioned and upcoming high-resolution near-infrared spectrographs and spectropolarimeters, such as GIANO \citep{Giano2006SPIE.6269E..19O}, CRIRES+ \citep{CRIRESPLUS2016SPIE.9908E..0ID}, IRD \citep{2012SPIE.8446E..1TT}, and SPIRou \citep{SPIROU2014SPIE.9147E..15A} will allow the investigation of the full near-IR wavelength domain and identification of better magnetic diagnostics for extended ranges of stellar effective temperatures and thus will potentially enable more systematic measurements of mean magnetic field strengths across the HR diagram, thereby covering a wider range of PMS stars.

\begin{acknowledgements}
O.K. acknowledges financial support from the Knut and Alice Wallenberg Foundation, the Swedish Research Council, and the Swedish National Space Board. G.J.H. is supported by general grant 11473005 awarded by the National Science Foundation of China.
Based on observations collected at the European Organisation for Astronomical Research in the Southern Hemisphere under ESO programmes 60.A-9051(A), 090.C-0131(B), and 69.C-0481(A).
Based on observations obtained at the Canada-France-Hawaii Telescope (CFHT), which is operated from the summit of Maunakea by the National Research Council of Canada, the Institut National des Sciences de l'Univers of the Centre National de la Recherche Scientifique of France, and the University of Hawaii. The observations at the Canada-France-Hawaii Telescope were performed with care and respect from the summit of Maunakea, which is a significant cultural and historic site. 
\end{acknowledgements}

\bibliographystyle{aa}
\bibliography{imtts}

\end{document}